\documentclass[sigconf]{acmart}
\AtBeginDocument{%
  }

\acmYear{2026}
\setcopyright{rightsretained}
\acmConference[GI '26]
  {Graphics Interface 2026}
  {June 9--12, 2026}
  {Waterloo, ON, Canada}
\acmBooktitle{Graphics Interface 2026 (GI '26),
  June 9--12, 2026, Waterloo, ON, Canada}
\acmDOI{}
\acmISBN{}

\usepackage{color, colortbl}
\usepackage{subfig}
\usepackage{multirow}
\usepackage{longtable}
\usepackage{hyperref}
\usepackage[font=small,labelfont=bf]{caption}
\usepackage{wasysym}
\usepackage[nolist,nohyperlinks, printonlyused, withpage]{acronym}
\usepackage{enumitem}

\begin{document}

\title{Proactive Systems in HCI and AI: Concepts, Challenges, and Opportunities}

\author{Nima Zargham}
\email{nima.zargham@utoronto.ca}
\orcid{0000-0003-4116-0601}
\affiliation{
  \institution{University of Toronto}
  \city{Toronto}
  \country{Canada}
}

\author{Sharon Ferguson}
\email{sharon.ferguson@uwaterloo.ca}
\orcid{0000-0002-2091-3435}
\affiliation{
  \institution{University of Waterloo}
  \city{Waterloo}
  \country{Canada}
}

\author{Jaisie Sin}
\email{JaisieSin@cunet.carleton.ca}
\orcid{0000-0001-6288-7303}
\affiliation{
  \institution{Carleton University}
  \city{Ottawa}
  \country{Canada}
}

\author{Cosmin Munteanu}
\email{cosmin@taglab.ca}
\orcid{0000-0002-0635-9124}
\affiliation{
  \institution{University of Waterloo}
  \city{Waterloo}
  \country{Canada}
}

\author{Anastasia Kuzminykh}
\email{anastasia.kuzminykh@utoronto.ca}
\orcid{0000-0002-5941-4641}
\affiliation{
  \institution{Faculty of Information \\ Department of Computer Science \\ University of Toronto}
  \city{Toronto}
  \country{Canada}
}
 
\renewcommand{\shortauthors}{Zargham et al.}

\begin{abstract}
The last few years have seen a significant rise in interest in highly autonomous and proactive systems, fueled by advances in AI. Systems that anticipate user needs, take initiative, and act without explicit user input. Such systems span a wide range of applications, from smart lighting that adapts to user activity to assistive robots that plan actions in advance to intelligent thermostats that learn routines and adjust environments proactively. Despite this breadth, the concept of proactivity remains loosely defined and inconsistently applied across research and practice. Current usage of the term often conflates fundamentally different system behaviors. For instance, simple reminders or recommendation systems are frequently labeled as proactive, even though underlying mechanisms and intentions differ significantly. This conceptual ambiguity limits our ability to systematically design, compare, and evaluate proactive systems. Moreover, existing methodologies for design and evaluation are largely rooted in reactive interaction paradigms, failing to address the unique challenges posed by proactive behavior, including timing, appropriateness, user control, transparency, and trust. This multidisciplinary workshop aims to establish a clearer and more rigorous foundation for understanding proactive systems. We bring together researchers and practitioners from Human-Computer Interaction, AI, and related fields to (1) develop a shared conceptualization of proactivity, (2) identify gaps and limitations in current design and evaluation approaches, and (3) co-create human-centered guidelines and research directions for future systems. Through interactive discussions and collaborative activities, the workshop seeks to map key challenges and opportunities, ultimately advancing robust and consistent frameworks for designing and evaluating proactive technologies.
\end{abstract}




\maketitle

\section{Motivation \& Goals}
In recent years, there has been a surge of interest in autonomous and proactive systems, driven by advances in artificial intelligence (AI), natural language processing (NLP), and sensing technologies. Unlike traditional reactive systems, proactive systems anticipate user needs, take initiative, and act without explicit instructions. Such systems span a wide range of applications, including smart homes \cite{Zargham2023Tickling, zargham2022Proactive}, assistive robotics \cite{Buyukgoz2021, Peng2019}, in-car assistants \cite{Kim2018}, health and wellbeing \cite{Lim2024, Ren2020}, and education \cite{Kraus2023, Kim2024}. For instance, navigation systems can suggest optimal routes before congestion occurs, fitness wearables can recommend activity adjustments based on user state, and productivity tools can anticipate scheduling conflicts or suggest task priorities. Emerging AI tools such as GitHub Copilot and Google Gemini further illustrate the growing role of proactive agents in both personal and professional contexts~\cite{Kraus2025IUI}. These systems present both exciting opportunities and significant challenges~\cite{berube2024proactive, oechsner2022challenges, zargham2022Proactive}. While the potential of proactive agents is vast, ranging from increasing perceived helpfulness to improving task efficiency and calibrating trust \cite{peng2019design, kraus2023improving, windl2024privacy}, there remains a critical gap in established guidelines and best practices for their human-centric design and evaluation. 
This starts with the description of the concept itself. Despite growing interest, the concept of proactivity remains loosely defined, conceptually imprecise, and inconsistently applied. Systems that send reminders or recommend content are often labeled proactive, even though their underlying mechanisms and intentions differ fundamentally from systems that genuinely anticipate and act on user goals. 
This ambiguity hampers our ability to systematically design, compare, and evaluate proactive systems. Existing design and evaluation methods are largely inherited from reactive paradigms and fail to capture critical aspects of proactive behavior, including timing, appropriateness, user control, transparency, and trust. From organizational psychology, proactivity is defined as self-initiated, future-oriented, and change-oriented behavior aimed at improving situations or preventing problems \cite{crant2000proactive, frese2001}. These definitions emphasize several key characteristics, including the anticipation of future states, goal-directed action, and persistence in shaping outcomes. Notably, proactivity in this tradition is not merely about acting independently, but about intentionally influencing future conditions.
However, research and practice often focus on superficial manifestations of system initiation, such as suggestions or prompts, rather than on the underlying capabilities that enable effective proactive behavior.

Moreover, current evaluation methods are primarily designed for reactive systems and fall short in addressing the unique demands of proactive system interactions. Commonly used tools such as the \textit{System Usability Scale} (SUS)~\cite{brooke2013sus} and the \textit{User Experience Questionnaire} (UEQ)~\cite{schrepp2017construction}, while often applied to proactive systems, have been validated only in reactive contexts. Consequently, they do not account for the nuances of proactive interactions, including autonomous decision-making and action execution. 
Aspects such as the appropriateness of interaction and the timing of proactive behavior initiation, the perceived alignment of system actions with user goals, and the user's sense of control and consent over unsolicited system actions are often overlooked.
Similarly, measures of social implications, such as perceived human-computer trust, rely on scales developed for reactive GUIs and are therefore of limited applicability~\cite{madsen2000measuring}.
For proactive systems, trust may also depend on factors such as the system's ability to justify its actions, predictability of autonomous behavior, and the extent to which users feel comfortable delegating initiative to the system.
The design of proactive systems also presents distinct challenges. Expectations of proactive behavior vary widely across application domains and user characteristics~\cite{meurisch2020exploring}. Users typically prefer to retain control, favoring initially low levels of proactivity that can increase with familiarity~\cite{glass2008toward}. While early design principles, building on mixed-initiative interaction~\cite{horvitz1999principles}, highlight qualities such as value, relevance, competence, controllability, transparency, and safety~\cite{yorke2012design}, existing human-computer interaction (HCI) guidelines only marginally address proactivity~\cite{amershi2019guidelines,Kraus2025IUI}. This points to a broader lack of standardized approaches for designing and evaluating proactive AI systems~\cite{berube2024proactive}.

This multidisciplinary workshop seeks to address these conceptual and methodological gaps by bringing together researchers and practitioners from HCI, AI, and related fields. Our workshop has the following objectives:
\begin{itemize}[leftmargin=*]
	\item Develop a shared conceptualization of what constitutes proactive behavior, and how it differs from related concepts such as autonomy, automation, and system-initiated interaction.
    \item Explore the design space of proactive systems to identify key dimensions, challenges, and contextual factors that shape how proactive behavior is designed, experienced, and perceived across domains.
    \item Establish principles, guidelines, and evaluation strategies that account for the unique challenges of proactive systems, including timing, transparency, user control, and trust.
\end{itemize}
Through collaborative discussions, scenario-based activities, and interactive knowledge sharing, the workshop aims to clarify the conceptual landscape of proactivity, highlight open research questions, and map practical opportunities for designing and evaluating proactive systems.


\section{Organisers}
The workshop is organised by leading researchers and practitioners from the fields of HCI and conversational user interfaces (CUIs). Multiple successful workshops were held between the organizers at HCI conferences, such as CHI~\cite{CUIatCHI2024}, HRI~\cite{McMillan2023,ringe2026HRI}, IUI~\cite{Kraus2025IUI}, IVA~\cite{IVA2025DEBP}, CUI~\cite{Avanesi2023Multimodal,voiceCraft2024,PersonasEvolved2025}, and MUM~\cite{Zargham2024Persona}. 

\noindent \textbf{Nima Zargham} is a postdoctoral researcher at the Faculty of Information at the University of Toronto. His research focuses on human-AI communication and collaboration. He has organized workshops at notable HCI conferences (e.g., CHI, CUI, HRI, IUI, and IVA) and serves on the steering committee of the ACM CUI conference series.

\noindent \textbf{Sharon Ferguson} is an Assistant Professor in Management Science and Engineering at the University of Waterloo. She leads the SHARE lab (Social and Human-centered AI for Reimagined Engagement), and her research focuses on AI and the future of work, particularly how AI can help and harm team collaboration. She has organized workshops at HCI and design conferences. 

\noindent \textbf{Jaisie Sin} is an Assistant Professor at Carleton University and the Canada Research Chair in Accessibility and Digital Technology. Her research investigates how to design technologies that are accessible, inclusive, and aligned with older adults’ needs. Her ongoing work investigates the development of conversational systems and agents that are proactive in supporting users often underrepresented in technology design.

\noindent \textbf{Cosmin Munteanu} is an Associate Professor and Schlegel Research Chair in Technology for Healthy Aging at the Department of Systems Design Engineering, University of Waterloo, and Director of the Technologies for Aging Gracefully lab. Their work is situated at the intersection of user experience design, digital inclusion, aging, natural language processing, and ethics, primarily focusing on the sociotechnical design of inclusive interfaces with and for older adults. Cosmin has previously organized several workshops in prestigious conferences including ACM CHI and ACM CUI.

\noindent \textbf{Anastasia Kuzminykh} is an Assistant Professor at the University of Toronto in the Faculty of Information, cross-appointed to the Department of Computer Science. She is an Associate Director of The Knowledge Media Design Institute, a member of The Data Science Institute, a Faculty Affiliate with the Schwartz Reisman Institute for Technology and Society, and the Director of the COoKIE AI research group (Communication, Organization of Knowledge, Information Ecosystems). Her work is situated at the intersection of AI Agent Design, Explainable AI, AI Safety, Natural Language Processing, and Conversational Interface Design. She has previously organized multiple workshops, including at the ACM CHI, ACM COMPASS, and IEEE CASCON.

\section{Schedule and Workshop Activities}
We aim to organize this workshop as a 3-hour event (two slots). The tentative schedule and activities are outlined below:  
\begin{itemize}[leftmargin=*]
    \item \textbf{Welcome \& Introduction (15 mins):} Brief introductions from the organizers and participants, outlining the workshop's theme and objectives, and providing an overview of the day's agenda. 
    \item \textbf{Activity 1 - Framing Proactivity (30 mins):} 
    Participants will be divided into small groups. Each group will discuss key characteristics of proactive systems and examine related terms such as autonomy, automation, and system-initiated interaction, identifying where these concepts overlap and where they diverge from proactivity. Each group will develop a brief working definition of proactivity, identify core dimensions, and provide examples or counterexamples. Insights will be shared in a brief plenary discussion to build a shared understanding and highlight open questions.
    \item \textbf{Presentation and Discussion (30 mins):} Workshop organizers will present a brief overview of prior research on proactive systems, highlighting key findings, design challenges, and open questions. Building on Activity 1, participants will reflect on how these research insights align with or challenge their previously developed definitions and dimensions of proactivity. A guided plenary discussion follows, in which we compare perspectives, surface similarities and differences, and collaboratively identify the key areas and dimensions most critical for designing and studying proactive systems. 
    \item \textbf{Coffee Break (30 mins):} Informal networking over coffee.
    \item \textbf{Activity 2 - Scenario Design (30 mins):} Participants will be divided into new small groups to develop scenarios involving people interacting with proactive systems. These can include both current and speculative future contexts. Each group will create a set of scenarios illustrating different forms of proactivity, focusing on user experience, system behavior, and potential challenges. 
    \item \textbf{Activity: Scenario Analysis (30 mins):} Groups will exchange scenarios with another group. For each scenario, they will identify key design considerations (e.g., timing, transparency, control, appropriateness) and propose potential evaluation approaches or metrics (e.g., trust, usefulness, user comfort). Insights will be briefly shared and discussed, synthesising insights across groups and linking them to broader workshop goals.
    \item \textbf{Closing (15 mins):} Organisers summarise the workshop's key outcomes. Together with participants, they identify concrete research questions, potential collaborations, and outline future directions for community-building.
\end{itemize}

\section{Format and Advertisement}
To reach further potential participants, we will distribute a Call for Participation through multiple channels, including social media platforms, mailing lists, and direct invitations to researchers with prior publications in related fields. The workshop will be held exclusively in person, with an expected attendance of 10-20 participants. Our dedicated workshop website (TBA) will provide detailed information on the call, key dates, and participation details. 

We will require a room equipped with a projector and, if available, a microphone. The space should have tables and chairs that can be arranged into small groups of 3-6 participants, as well as empty wall space for attaching poster boards for group activities. The organizers will provide detailed setup requirements for all planned activities in advance. Attendance in this workshop will be open to GI participants without requiring paper submissions.

\paragraph{Inclusion \& Accessibility}
We are committed to fostering an inclusive and accessible environment for all attendees and will work closely with the conference organizers to ensure this. We will proactively address participants' identified accessibility needs and provide the necessary resources to support their engagement. 
Drawing from our experience organizing similar events, we strive to create a space where all participants are valued as equal contributors to the discussion. We will further use online tools like Google Docs to record our workshop contributions in an accessible way.
Additionally, we aim to encourage meaningful connections by facilitating structured breakout sessions and mentorship opportunities and create spaces for cross-disciplinary collaboration and learning.

\section{Post-Workshop Plans}
Key discussion points and insights will be documented on an open online platform during and after the workshop to enable continued engagement. Participants will also be encouraged to explore future collaborative projects inspired by the discussions. To sustain community engagement beyond the workshop, we will create a dedicated Discord channel to keep participants connected and informed about developments in the topic.

\bibliographystyle{ACM-Reference-Format}
\bibliography{sample-base}

\end{document}